\documentclass[11pt]{article}
\usepackage{latexsym}  % for Symbol fonts
\usepackage{amssymb}
\usepackage{graphicx}
\usepackage{amsmath}
\begin{document}

\begin{center}
{\Large\bf How to Evade a No-Go Theorem in Flavor Symmetries
}\footnote{
Invited talk given at {\it International Workshop on Grand Unified 
Theories: Current Status and Future Prospects}, Ritsumeikan University, 
Kusatsu, Shiga, Japan, December 17 - 19, 2007. \\
To be published in Proceedings.
}

%\classification{11.30.Hv, 12.15.Ff, 12.90.+b}
%\keywords      {flavor symmetry, quark and lepton masses and mixings}

\vspace{3mm}
{\bf Yoshio Koide}

{\it IHERP, Osaka University, 1-16 Machikaneyama, 
Toyonaka, Osaka 560-0043, Japan 
}

{\it koide@het.phys.sci.osaka-u.ac.jp
}

\end{center}

\vspace{3mm}
\begin{abstract}
A no-go theorem in flavor symmetries is reviewed.
The theorem asserts that we cannot bring any flavor symmetry into
mass matrix model in which number of Higgs scalars is, at most,
one for each sector (e.g. $H_u$ and $H_d$ for up- and down-quark
sectors, respectively). Such the strong constraint comes from 
the SU(2)$_L$ symmetry.  Possible three options to evade the 
theorem are discussed.
\end{abstract}

%\maketitle

%%%%%%%%%%%%%%%%%%%%%%%%%%%%%%%%%%%%%%%%%%%%
%% MAINMATTER
%%%%%%%%%%%%%%%%%%%%%%%%%%%%%%%%%%%%%%%%%%%%

\section{Introduction}

When we see a history of physics, we will find that ``symmetries" always 
play a key role in the new physics.  
For investigating an origin of flavors, 
too, we may expect that an approach based on symmetries will be a powerful
instrument. 
Especially, how to treat the flavor symmetry is a big concern in 
grand unification model-building.

However, when we want to introduce a flavor symmetry (e.g. discrete one, U(1), 
and so on) into our mass matrix model, we always encounter an obstacle, 
a no-go theorem in flavor symmetries \cite{no-go}.
The theorem asserts that 
we cannot bring any flavor symmetry into a mass matrix model
in which number of Higgs scalars is, at most, one for each sector 
(e.g. $H_u$ and $H_d$ for up- and down-quark sectors, respectively). 
Such the strong constraint comes from the SU(2)$_L$ symmetry.
We must take this theorem seriously. 
This theorem seems to require a new idea for the mass generation
against a conventional idea ``(masses)=(Yukawa coupling constants)
$\times$(vacuum expectation value of Higgs scalar)".  
We should not consider this theorem to be negative, and 
we should utilize this theorem positively to investigate the origin 
of the flavor mass spectra.

Nevertheless, there are some optimists.
Why?
They know that, for example, the Yukawa
interactions in the up- and down-quark sectors are independent
of each other, and,
besides, the Higgs scalars which contribute to each sector
can be different (e.g. $H_u$ and $H_d$, respectively).
Therefore, they consider that we can apply the flavor 
symmetry to the up- and down-quark sectors separately.
First, let check this. 

In the standard model, the fermion masses are generated from 
the vacuum expectation values (VEVs) of the Higgs scalars:
$$
H_{Y}= (Y_u)_{ij} \overline{Q}_{Li} H_u u_{Rj} 
+ (Y_d)_{ij} \overline{Q}_{Li} H_d d_{Rj} +h.c.
\eqno(1.1)
$$ 
where
$$
Q_L = \left(
\begin{array}{c}
u_L \\
d_L
\end{array} \right) , \ \ 
H_u = \left(
\begin{array}{c}
H_u^0 \\
H_u^-
\end{array} \right) , \ \ 
H_d = \left(
\begin{array}{c}
H_d^+ \\
H_d^0
\end{array} \right) . 
\eqno(1.2)
$$ 
Therefore, the mass matrices $M_u$ and $M_d$ for the up- and
down-quarks are given by
$$
(M_u)_{ij} = (Y_u)_{ij} \langle H_u^0 \rangle , \ \ 
(M_d)_{ij} = (Y_d)_{ij} \langle H_d^0 \rangle .
\eqno(1.3)
$$  
The requirement of a flavor symmetry means that
the interactions (1.1) are invariant
under the transformation of the flavor basis
$$
\begin{array}{lll}
Q_L & \rightarrow & Q'_L=T_L Q_L ,\\
u_R & \rightarrow & u'_R =T_R^u u_R ,\\
d_R & \rightarrow & d'_R =T_R^d d_R .
\end{array}
\eqno(1.4)
$$
Then, the requirement, in general, imposes the following 
constraints on the Yukawa coupling constants $Y_u$ and $Y_d$:
$$
T_L^\dagger Y_u T_R^u = Y_u , \ \ 
T_L^\dagger Y_d T_R^d = Y_d .
\eqno(1.5)
$$
(Of course, the physical quark masses (1.3) are given below the
energy scale $\mu=\Lambda_{EW}$, at which the SU(2)$_L$ symmetry
is broken, so that the constraint (1.5) has a meaning above
$\mu=\Lambda_{EW}$, i.e. for $Y_u(\mu)$ and $Y_d(\mu)$ at
$\mu > \Lambda_{EW}$.)
This constraint (1.5) does not always mean that the form of $Y_u$ is
the same as that of $Y_d$.
A relation between the coupling constants $Y_u$ and $Y_d$ looks 
like free.

However, when we take notice of the Hermitian matrices 
$Y_u Y_u^\dagger$ and $Y_d Y_d^\dagger$, 
the situation will become clear:
$$
\begin{array}{l}
T_L^\dagger Y_u (Y_u)^\dagger T_L = Y_u (Y_u)^\dagger , \\ 
T_L^\dagger Y_d (Y_d)^\dagger T_L = Y_d (Y_d)^\dagger .
\end{array}
\eqno(1.6)
$$ 
Here, we should note that the flavor transformation 
operator $T_L$ is identical both for up- and down-quark
sectors.
As we discuss in the next section, this will give a strong 
constraint for the Cabibbo-Kobayashi-Maskawa \cite{CKM} (CKM) 
mixing matrix $V$, which is defined by
$$
 V = (U_L^u)^\dagger U_L^d ,
\eqno(1.7)
$$
where $U_L^f$ are defined by
$$
\begin{array}{l}
(U_L^u)^\dagger M_u U_R^u = D_u \equiv {\rm diag}(m_u,m_c,m_t) ,\\[.1in]
(U_L^d)^\dagger M_d U_R^d = D_d \equiv {\rm diag}(m_d,m_s,m_b) ,
\end{array}
\eqno(1.8)
$$
i.e.
$$
\begin{array}{l}
(U_L^u)^\dagger Y_u (Y_u)^\dagger U_L^u = \frac{1}{v_u^2} D_u (D_u)^\dagger  ,\\ 
(U_L^d)^\dagger Y_d (Y_d)^\dagger U_L^d = \frac{1}{v_d^2}D_d (D_d)^\dagger  ,
\end{array}
\eqno(1.9)
$$ 
and $v_u=\langle H_u^0\rangle$ and $v_d=\langle H_d^0\rangle$.
  
%%%%%%%%%%%%%%%%%%%%%%%%%%%%%%%%%%%%%%%%%%%%%%%%%%%%%%%%%%%%
\section{No-Go theorem in flavor symmetries}

The no-go theorem in flavor symmetries is as follows \cite{no-go}:

{\bf [Theorem] }  When a flavor symmetry is brought into a model 
within the framework of the standard model, the flavor mixing matrix 
(CKM matrix and/or neutrino mixing matrix) cannot describe 
a mixing among 3 families, and only a mixing between 2 families 
is allowed. 

For example, the theorem asserts that we can obtain only
a two-flavor mixing such as 
$$
V= \left(
\begin{array}{ccc}
\ast & \ast & 0 \\
\ast & \ast & 0 \\
0 & 0 & 1 
\end{array} \right).
\eqno(2.1)
$$

Such a strong constraint comes from the relations 
(1.6) and (1.9).
When we define the following operators 
$$
T_u =(U_L^u)^\dagger T_L U_L^u , \ \ 
T_d =(U_L^d)^\dagger T_L U_L^d ,
\eqno(2.2)
$$
from the  flavor transformation operator $T_L$ in the flavor symmetry,
we can obtain a relation
$$
T_u^\dagger D_u^2 T_u =D_u^2 , \ \ 
T_d^\dagger D_d^2 T_d =D_d^2 ,
\eqno(2.3)
$$
because $Y_f Y_f^\dagger$ in (1.9) is express as
$$
Y_f Y_f^\dagger = T_L^\dagger Y_f Y_f^\dagger T_L 
= T_L^\dagger \cdot U_L^f (1/v_f^2) D_f D_f^\dagger 
(U_L^f)^\dagger T_L 
=U_L^f T_f^\dagger (1/v_v^2) D_f D_f^\dagger T_f (U_L^f)^\dagger .
\eqno(2.4)
$$                                                          
Therefore, if the eigenvalues of $Y_{f}$ are non-zero and non-degenerate, 
the operator $T_f$ must be expressed by a form of the phase matrix 
$$
T_f=P_f \equiv {\rm diag}(e^{i \delta_1^f},
e^{i \delta_2^f},e^{i \delta_3^f}) ,
\eqno(2.5)
$$
so that $T_L$ is expressed as
$$
T_L=U_L^u P_u (U_L^u)^\dagger =U_L^d P_d (U_L^d)^\dagger ,
\eqno(2.6)
$$
from the definition of $T_f$, Eq.(2.2).
Therefore, the phase matrices $P_u$ and $P_d$ are related as
$$
P_u =(U_L^u)^\dagger U_L^d P_d (U_L^d)^\dagger U_L^u
=V P_d V^\dagger ,
\eqno(2.7)
$$
from Eq.(2.6) and the definition of the CKM matrix $V$, (1.7).
Thus, we obtain a constraint on the CKM matrix $V$
$$
P_u V -V P_d =0 ,
\eqno(2.8)
$$
i.e.
$$
 (e^{i\delta_i^u} -e^{i\delta_j^d})
V_{ij} =0 .
\eqno(2.9)
$$
Therefore, if $\delta_i^u \neq \delta_j^d$, we obtain an 
unwelcome result $V_{ij} = 0$ \cite{no-go}.

We do not consider the case with $\delta_1^u=\delta_2^u=\delta_3^u$
and $\delta_1^d=\delta_2^d=\delta_3^d$ , because the case corresponds 
to a trivial flavor transformation $T_L ={\bf 1}$.
For a non-trivial flavor transformation $T_L$, we must choose, at 
least, one of $\delta_i^f$ differently from others.   
For example, for the case with $\delta_1^f=\delta_2^f \neq \delta_3^f$, 
we can obtain only a two-family mixing 
$$
V= \left(
\begin{array}{ccc}
\ast & \ast & 0 \\
\ast & \ast & 0 \\
0 & 0 & 1 
\end{array} \right) .
\eqno(2.10)
$$

We can essentially obtain a similar result in the lepton sectors
\cite{no-go}, 
although a stronger constraint will be added if the neutrino mass matrix 
is Majorana type. 

Now, let us summarize the premises to derive the theorem:

(i) The SU(2)$_L$ symmetry is unbroken;

(ii) There is only one Higgs scalar in each sector;

(iii) 3 eigenvalues of $Y_f$ in each sector are non-zero and no-degenerate.

\noindent If one of them in a model is not satisfied, the model can evade 
the theorem.

For example, let us consider a model:  
(i) we consider an unbroken flavor symmetry at GUT scale; 
(ii) there is only one Higgs scalar in each sector, e.g. $H_u$ 
and $H_d$; 
(iii)  3 eigenvalues of $Y_f$ in each sector are completely 
different from each other and not zero at the GUT scale.
Then, such the model is ruled out by the theorem.
However, if the flavor symmetry is explicitly broken, i.e.
the model has an explicit symmetry breaking term at the
beginning,  
the present theorem does not affect such a model.

%%%%%%%%%%%%%%%%%%%%%%%%%%%%%%%%%%%%%%%%%

\section{How to evade  the no-go theorem}

We will discuss three options to evade the no-go theorem:

(A) Model with multi-Higgs scalars;

(B) Model with an explicit broken flavor symmetry;

(C) Model in which $Y$'s are fields.

\noindent Of course, a model in which all SU(2)$_L$ doublets
are singlets (i.e. $T_L={\bf 1}$) under the flavor symmetry 
can evade the no-go theorem.

In most phenomenological studies of flavor symmetries,
models contain a phenomenological symmetry breaking
term from the beginning, although we suppose that 
such a symmetry breaking term is spontaneously generated 
from the world of an unbroken flavor symmetry. 
Such a model belongs to the category (B).
In the present studies, we do not discuss such a model
in which the problem is postponed to future.  

In most GUT models, more than two Higgs scalars which
belong to different multiplets of the GUT group are
assumed.
Such models belong to the category (A).
In such models, it is essential whether unwelcome components of 
the Higgs scalars can naturally be suppressed in the Higgs potential
without any explicit symmetry breaking term.

%%%%%%%%%%%%%%%%%%%%%%%%%
\subsection{Model with multi-Higgs scalars}

A model with multi-Higgs scalars can evade the no-go 
theorem, where the Higgs scalars must have different transformation 
properties under the flavor symmetry.
For example, we may consider a model:
$$
M_{ij} = Y_{ij}^a \langle H_a \rangle +
Y_{ij}^b \langle H_b \rangle +
Y_{ij}^c \langle H_c \rangle .
\eqno(3.1)
$$
However, generally, such a multi-Higgs model induces the so-called 
flavor-changing neutral currents (FCNC) problem.
We must make those Higgs scalars heavy except for one of linear 
combinations of those scalars, e.g.
$$
U_H \left(
\begin{array}{c}
H_a \\
H_b \\
H_c 
\end{array} \right) = \left(
\begin{array}{c}
H_0 \\
H_1 \\
H_2 
\end{array} \right)
\begin{array}{c}
\sim 10^2 {\rm GeV} \\
\sim 10^{16} {\rm GeV} \\
\sim 10^{16} {\rm GeV} ,
\end{array}
\eqno(3.2)
$$
where $U_H$ is a mixing matrix among $H_a$, $H_b$ and $H_c$. 
Since the Higgs scalars $H_a$ $H_b$ and $H_c$ have different
quantum numbers of the flavor symmetry, such a mixing (3.2)
breaks the flavor symmetry at a high energy scale, at which
the mixing $U_H$ is caused. 
Of course, such a mixing must be caused without any explicit
symmetry breaking parameters. 

However, at present, models which give a reasonable mixing 
mechanism are few.  
The mechanism must be proposed in the framework of the exact flavor 
symmetry. In most models, the suppression of unwelcome 
components are only assumptions by hand.  

%%%%%%%%%%%%%%%%%%%%%%%%%%%

\subsection{Model with an explicitly broken symmetry}

We consider a model in which the symmetry is explicitly broken from 
the beginning. 
In other words, in such a model, there is no flavor symmetry from
the beginning.
Therefore, such a model can, of course, evade the no-go theorem.

In most phenomenological studies of flavor symmetries,
models contain a phenomenological symmetry breaking
term from the beginning, although we suppose that 
such a symmetry breaking term is spontaneously generated 
from the world of an unbroken flavor symmetry. 
Such a model belongs to the present category.

In any flavor symmetry, the symmetry finally has to
be broken badly, because the observed flavor mass values are
highly hierarchical.
As an example of a model 
in which the flavor symmetry is badly broken at the beginning,
let us review the following model \cite{Haba-Koide07}: 
We assume a U(3) flavor symmetry. 
For simplicity, we consider only a case of the charged lepton sector. 
The symmetry U(3) is broken by parameters $(Y_e)_{ij}$ explicitly:
$$
W_Y = \sum_{i,j} (Y_e)_{ij} L_j E_i H_d.
\eqno(3.3)
$$
(For convenience, hereafter, we will drop the index ``e".)
Also, we consider a U(3) nonet field $\Phi$    
and we denote the superpotential for $\Phi$  as
$$
W_\Phi = m_1 {\rm Tr}[\Phi\Phi] + m_2 {\rm Tr}^2[\Phi]
+\lambda_1 {\rm Tr}[\Phi\Phi\Phi]
+\lambda_2 {\rm Tr}[\Phi\Phi] {\rm Tr}[\Phi] +
\lambda_3 {\rm Tr}^3[\Phi] .
\eqno(3.4)
$$
We assume that the symmetry is also broken by a tadpole term with 
the same symmetry breaking parameter $Y$ as follows:
$$
W= W_\Phi -\mu^2 {\rm Tr}[Y \Phi] + W_Y .
\eqno(3.5)
$$
Then, we obtain
$$
\frac{\partial W}{\partial \Phi} = 0 =\frac{\partial W_\Phi}{\partial \Phi}
-\mu^2 Y ,
\eqno(3.6)
$$
where
$$
\frac{\partial W_\Phi}{\partial \Phi}
=
 3 \lambda_1 \Phi \Phi
 + c_1 (\Phi) \Phi + c_0(\Phi) {\bf 1} ,
\eqno(3.7)
$$
$$
c_1(\Phi)= 2(m_1 +\lambda_2 {\rm Tr}[\Phi] ) ,
\eqno(3.8)
$$
$$
c_0(\Phi)=2 m_2 {\rm Tr}[\Phi] +
\lambda_2 {\rm Tr}[\Phi\Phi] +3\lambda_3 {\rm Tr}^2[\Phi] .
\eqno(3.9)
$$

Now, we put an ansatz that our vacuum is given by the following
specific solution of 
Eq.(3.6):
$$
3 \lambda_1 \Phi \Phi -\mu^2 Y =0 ,
\eqno(3.10)
$$
and
$$
c_1 (\Phi) \Phi + c_0(\Phi) {\bf 1} =0 .
\eqno(3.11)
$$
Eq.(3.10) leads to a bilinear mass formula
$$
Y_{ij} = \frac{3\lambda_1}{\mu^2} \sum_k \langle\Phi_{ik}\rangle 
\langle\Phi_{kj}\rangle .
\eqno(3.12)
$$
For non-zero and non-degenerate eigenvalues $v_i$ of
$\langle \Phi\rangle$, Eq.(3.11) requires $c_1=0$ and 
$c_0=0$.
Thus, we can obtain a relation for the charged lepton masses
by choosing a suitable form of $W_\Phi$.

For example, when we assume \cite{Haba-Koide07}
$$
W_\Phi = m {\rm Tr}[\Phi\Phi] 
+ \lambda {\rm Tr}[\Phi^{(8)}\Phi^{(8)}\Phi^{(8)}] ,
\eqno(3.13)
$$
where $\Phi^{(8)}$ is an octet part of the nonet scalar $\Phi$:
$$
\Phi^{(8)}=\Phi -\frac{1}{3}{\rm Tr}[\Phi] {\bf 1} ,
\eqno(3.14)
$$
we obtain
$$
{\rm Tr}[\Phi\Phi] = \frac{2}{3}{\rm Tr}^2[\Phi] ,
\eqno(3.15)
$$
from $c_0=0$, because
$$
{\rm Tr}[\Phi^{(8)}\Phi^{(8)}\Phi^{(8)}]
={\rm Tr}[\Phi\Phi\Phi]-{\rm Tr}[\Phi] \left({\rm Tr}[\Phi\Phi]-
\frac{2}{9}{\rm Tr}^2[\Phi]\right) .
\eqno(3.16)
$$
Eq.(3.15) leads to the VEV relation
$$
v_1^2+v_2^2+v_3^2 =\frac{2}{3}(v_1+v_2+v_3)^2 ,
\eqno(3.17)
$$
in the diagonal basis of $\langle\Phi_{ij}\rangle =\delta_{ij} v_i$.
Therefore, from Eqs.(3.12) and (3.17), we obtain the charged lepton 
mass formula \cite{Koidemass}
$$
m_e+m_\mu+m_\tau=\frac{2}{3}(\sqrt{m_e}+\sqrt{m_\mu}+\sqrt{m_\tau})^2 .
\eqno(3.18)
$$
The formula (3.18) can give an excellent prediction $m_\tau = 1776.97$ 
MeV from the observed values of $m_e$ and $m_\mu$, which is in
excellent agreement with the observed value \cite{PDG06}  $m_\tau^{obs}=1776.99^{+0.29}_{-0.26}$ MeV.

%%%%%%%%%%%%%%%%%%%%%%%%%%%%%%%%%%%%%%%%%%%%%%%%%%%%%%%%%%%%%%%%%%%%%%%%

\subsection{Model in which $Y$'s are fields}

We consider that $Y_f$ in the Yukawa interaction (1.1) are 
fields, e.g. 
$$
H_{Y}= \frac{(Y_u)_{ij} }{M}\overline{Q}_{Li} H_u u_{Rj} 
+ \frac{(Y_d)_{ij}}{M} \overline{Q}_{Li} H_d d_{Rj} +h.c.
\eqno(3.19)
$$ 
Since the fields $Y_f$ are transformed as 
$$
 Y_f \rightarrow Y'_f = T_L Y_f (T_R^f)^\dagger ,
\eqno(3.20)
$$
under the transformation (1.4), 
the constraints (1.6) for $Y_f (Y_f)^\dagger$ disappears, 
so that we can again evade the no-go theorem.
                                                                   
For example, recently, Haba \cite{Haba} has suggested that the effective Yukawa interaction originates in a higher dimensional term in K\"{a}hler 
potential $K$
$$
K \sim \frac{1}{M^2} y_A A_{ij}^\dagger L_j E_i H_d ,
\eqno(3.21)
$$
which leads to an effective Yukawa interaction
$$
(K)_D \sim \frac{1}{M^2} y_A (F_A^\dagger)_{ij} L_j E_i H_d.
\eqno(3.22)
$$
A similar idea in the neutrino masses has been proposed by Arkani-Hamed, Hall, Murayama, Smith and Weiner \cite{Murayama01}.

For example, when we adopt an O'Raifeartaigh-type SUSY breaking mechanism 
\cite{ORft}
$$
W=W_\Phi(\Phi) +\lambda_A {\rm Tr}[A\Phi\Phi]
+\lambda_B {\rm Tr}[B\Phi\Phi] -\mu^2 {\rm Tr}[\xi A] ,
\eqno(3.23)
$$
where  $\xi$  ($3\times 3$ matrix) is a flavor breaking parameter,
the scalar potential $V$ is given by 
$$
V= {\rm Tr}\left[ (\lambda_A \Phi\Phi-\mu^2 \xi)
(\lambda_A \Phi\Phi-\mu^2 \xi)^\dagger\right]
+|\lambda_B|^2 {\rm Tr}[\Phi\Phi\Phi^\dagger \Phi^\dagger]
$$
$$
+{\rm Tr}\left[ \left( \frac{\partial W_\Phi}{\partial \Phi} 
+(\lambda_A A +\lambda_B B)\Phi+\Phi(\lambda_A A +\lambda_B B) \right)
\left( \frac{\partial W_\Phi}{\partial \Phi} 
+(\lambda_A A +\lambda_B B)\Phi+\Phi(\lambda_A A +\lambda_B B) \right)^\dagger
\right] ,
\eqno(3.24)
$$
so that the conditions $\partial V/\partial A=0$ and 
$\partial V/\partial B=0$ give the constraint  
$$
 \frac{\partial W_\Phi}{\partial \Phi} + (\lambda_A A+ \lambda_B B)\Phi 
+\Phi (\lambda_A A+ \lambda_B B) =0 ,
\eqno(3.25)
$$
and the condition $\partial V/\partial \Phi=0$ gives 
$$
(|\lambda_A|^2+|\lambda_B|^2) \Phi\Phi =  \lambda_A^* \mu^2 \xi , 
\eqno(3.26)
$$
under the condition (3.25).
Therefore, we can again obtain a bilinear form for the effective 
Yukawa coupling constant as follows:
$$
-F_A^\dagger  = \frac{\partial W}{\partial A} 
 =\lambda_A \Phi\Phi-\mu^2 \xi
=\lambda_B\frac{\lambda_B}{\lambda_A} \Phi\Phi \neq 0 ,
\eqno(3.27)
$$
$$
-F_B^\dagger =\frac{\partial W}{\partial B}=
\lambda_B \Phi\Phi \neq 0 .
\eqno(3.28)
$$
However, since the present model leads to unwelcome situation that
fermion parts $\psi_{C'}$ of the superfields 
$C'\equiv (\lambda_B A -\lambda_A B)/\sqrt{\lambda_A^2+\lambda_B^2}$
become massless.
In order to make those massless fermions harmless, we must
change the K\"{a}helar potential $K$ into a non-canonical form
with higher dimensional terms
$$
- \frac{1}{M^2} \left({\rm Tr}^2 [A^\dagger A] + 
{\rm Tr}^2[B^\dagger B] \right) .
\eqno(3.29)
$$
Two conditions $\partial V/\partial A=0$ and 
$\partial V/\partial B=0$ have led to the same constraint (3.25) 
in the canonical K\"{a}heler potential,
while, in the non-canonical  K\"{a}heler potential with the higher 
dimensional terms (3.29), two conditions $\partial V/\partial A=0$ and 
$\partial V/\partial B=0$ lead to different constraints, and thereby,
we can obtain
$$
\langle A\rangle = \langle B\rangle =0\ \  {\rm and} \ \ 
\langle \frac{\partial W_\Phi}{\partial \Phi} \rangle = 0 .
\eqno(3.30)
$$
Therefore, the massless fermions $\psi_{C'}$ can become harmless,
because some dangerous amplitudes become zero due to 
$\langle A\rangle = \langle B\rangle =0$.
On the other hand, the VEV spectrum of $\Phi$ is practically determined 
by the constraint 
$$
\frac{\partial W_\Phi}{\partial \Phi}=0 ,
\eqno(3.31)
$$
which is derived from the conditions $\partial V/\partial A=0$ and 
$\partial V/\partial B=0$. 
By assuming a suitable form of $W_\Phi$, we can again obtain the mass 
relation (3.18).
For more details, see Ref.\cite{Haba-Koide08}.

%%%%%%%%%%%%%%%%%%%%%%%%%%%%%%%%%%%%%%%%%%%%%%%%%%%%%%%%
\section{Summary}

The no-go theorem in  flavor symmetries asserts us that we cannot bring 
any flavor symmetry into a mass matrix model based on 
the standard model.
We have demonstrated three options to evade the no-go theorem 
in the flavor symmetries:

(A) Model with multi-Higgs scalars;

(B) Model with an explicit broken symmetry;

(C) Model in which $Y$'s are fields.

Models based on the scenario (A) have been proposed by many authors.
In most GUT models, more than two Higgs scalars which
belong to different multiplets of the GUT group are assumed.
Therefore, if we can make those scalars heavy except one component,
we will obtain a reasonable model which can evade the no-go theory.   
However, current most models have not demonstrated an explicit  
mechanism (Higgs potential) which makes unwelcome components of the 
Higgs scalars heavy except for one.

In most phenomenological studies of flavor symmetries,
models contain a phenomenological symmetry breaking
term from the beginning.
Even if we suppose that 
such a symmetry breaking term is spontaneously generated 
from the world of an unbroken flavor symmetry,
the model belongs to the category (B), unless we explicitly 
demonstrate it on the basis of a Higgs potential
without any symmetry breaking term.
In the scenario (B), there is no flavor symmetry from the beginning.  
The ``flavor symmetry" is a faked one for convenience.  
However, if we once suppose a flavor symmetry, rather, 
we would like to consider that the symmetry is exact, and then 
it is broken spontaneously.  Therefore, we are still unsatisfactory 
to the scenario (B).

Models based on the scenario (C) are interesting.  
However, in order to give an effective Yukawa interaction, 
we need a term with higher dimension 
$$
\frac{1}{M} (Y_e)_{ij} L_j E_i H_d
$$
in the superpotential $W$, or
$$
 \frac{1}{M^2} (Y_e^\dagger)_{ij} L_j E_i H_d
$$
in the K\"{a}hler potential $K$.
However, we want a model without such higher dimensional terms 
as possible.

In conclusion, we have proposed three options to evade the no-go theorem.   
Those scenarios can evade the no-go theorem practically, but 
those still do not suit our feeling.   
We must seek for a more natural scenario  which is free 
from the no-go theorem.
Then, a hybrid model between (A) and (C), where there is a U(3)-flavor 
nonet Higgs doublet scalar and only one component becomes light, 
will be promising.

%%%%%%%%%%%%%%%%%%%%%%%%%%%%%%%%%%%%%%%%%%%%%%%%
%% BACKMATTER
%%%%%%%%%%%%%%%%%%%%%%%%%%%%%%%%%%%%%%%%%%%%%%%%

{\large\bf Acknowledgments}

  The author would like to thank Professor T.~Fukuyama for
giving an opportunity for talking in this workshop.  
The work is supported by the Grant-in-Aid for
Scientific Research, Ministry of Education, Science and 
Culture, Japan (No.18540284).

%%%%%%%%%%%%%%%%%%%%%%%%%%%%%%%%%%%%%%%%%%%

\end{document}